# Schemes for Privacy Data Destruction in a NAND Flash Memory

**Na-Young Ahn[1] and Dong Hoon Lee[2]**
[1]The Graduate School of Information Security at Korea University, Seoul, Korea
[2]CIST & The Graduate School of Information Security at Korea University, Seoul, Korea

Corresponding author: Dong Hoon Lee (e-mail: donghlee@ korea.ac.kr).

**ABSTRACT** We propose schemes for efficiently destroying privacy data in a NAND flash memory. Generally, even if privcy data is discarded from NAND flash memories, there is a high probability that the data will remain in an invalid block. This is a management problem that arises from the specificity of a program operation and an erase operation of NAND flash memories. When updating pages or performing a garbage collection, there is a problem that valid data remains in at least one unmapped memory block. Is it possible to apply the obligation to delete privacy data from existing NAND flash memory? This paper is the answer to this question. We propose a partial overwriting scheme, a SLC programming scheme, and a deletion duty pulse application scheme for invalid pages to effectively solve privacy data destruction issues due to the remaining data. Such privacy data destruction schemes basically utilize at least one state in which data can be written to the programmed cells based on a multi-level cell program operation. Our privacy data destruction schemes have advantages in terms of block management as compared with conventional erase schemes, and are very economical in terms of time and cost. The proposed privacy data destruction schemes can be easily applied to many storage devices and data centers using NAND flash memories.

## I. INTRODUCTION

Nonvolatile memories are divided into a memory which is overwritable and a memory which is not. NAND flash memory is a memory that cannot be overwritten [1,2]. NAND flash memory basically adjusts the threshold voltage by moving the trap charge according to the applied voltage. Data corresponding to the adjusted threshold voltage is determined. Also, in terms of management, NAND flash memory has an issue about residual data related to deletion because the writing unit and the erasing unit are different. In general, valid data is collected into one block through a garbage collection, and the remaining invalid data is erased in units of blocks when a certain standard is reached internally [3-5]. However, due to the characteristics of NAND flash memory, NAND flash memory can easily obtain meaningful information through digital forensics even if the user deletes the data [6,7]. If such meaningful information is obtained by investigative agencies such as police/prosecutors, nothing better. But what about the opposite? Because digital forensics is easy, anyone can illegally obtain information related to the user through digital forensics. If this information is privacy data, the problem is obvious.

Existing NAND flash memories cannot be overwritten. Therefore, even if the user deletes privacy data stored in NAND flash memory, the probability that the original type of data is still left is considerable. Recently, data centers are using SSDs that use NAND flash memories, which is likely to leave unprotected privacy data on SSDs in the data center [8,9]. Naturally, the duty to delete privacy data also applies to NAND flash memory. It should not be considered technically impossible or free from these obligations. In fact, technically speaking, even if you don't change the structural parts of existing NAND flash memory, you don't have to apply a lot of obligations to delete them when you use partial program operations. Ahn and Lee have already begun discussions to solve these problems [9], and in this paper we deepen it. They introduced the destruction of privacy data (personal information) using an overwritable program operation rather than an erase operation. In 2018, Lin et al. presented a simple way to sanitize MLC data by one shot programming [10]. Lin's one shot programming scheme, however, has the problem of varying the storage capacity of NAND flash memory. Since Lin's scheme changes from the number of MLC pages to SLC pages, the number of pages in one memory block suddenly decreases. This means changing the size of the storage capacity. Thus, this approach is unlikely to be used from a management point of view.

The duty to delete privacy data from NAND flash memories should be applied in the near future. There are some limitations to this in existing technology, but we are challenging these limitations by using algorithms that apply a simple deletion duty.

The structure of our paper is as follows. In Section II, we explored why privacy data inevitably remains in NAND flash memories and how we can handle the remaining privacy data. In Section III, we confirmed that there are overwritable states in NAND flash memory and that these states can be overwritten in NAND flash memory. In section IV, we proposed a scheme of destroying privacy data using partial overwriting and a scheme of destroying privacy data using a deletion duty pulse. In section V, we compared the performances with respect to processing time and cell degradation characteristics between the conventional block deletion schemes and the proposed deletion schemes.

The proposed data destruction schemes can be immediately applied to existing NAND flash memory and allow simple and quick destruction of privacy data without affecting the cell characteristics. In the future, our data destruction schemes are expected to make a significant contribution to NAND flash memory, which must be subject to the duty to delete.

## II. Related Works for NAND Flash Memory

In general, a privacy data life cycle consists of collection, storage, storage, use / provision, and destruction. Privacy data destruction is the final stage of the privacy data life cycle. As rights as the subject of information become more prominent, the importance of privacy data destruction is increasingly emphasized. Recently, Europe Unites have legally obliged to destroy privacy data in terms of enhancing rights as an individual's information subject [8]. NIST has suggested guidelines for media sanitization. Media sanitization is largely divided into Clear/Purge/Destroy [11].

'Clear' is a technique for overwriting non-sensitive data using software or hardware. This overwrites the new value on the storage device using a normal read or write command. In the case of a device that is not provided with overwriting, such as a NAND flash memory, clear means that it is set to a factory state [11]. 'Purge' is a physical or logical technique that uses lab schemes to make target data unrecoverable [11]. 'Destroy' means the use of laboratory technology to render target data unrecoverable while at the same time not functioning as a medium for storing the data.

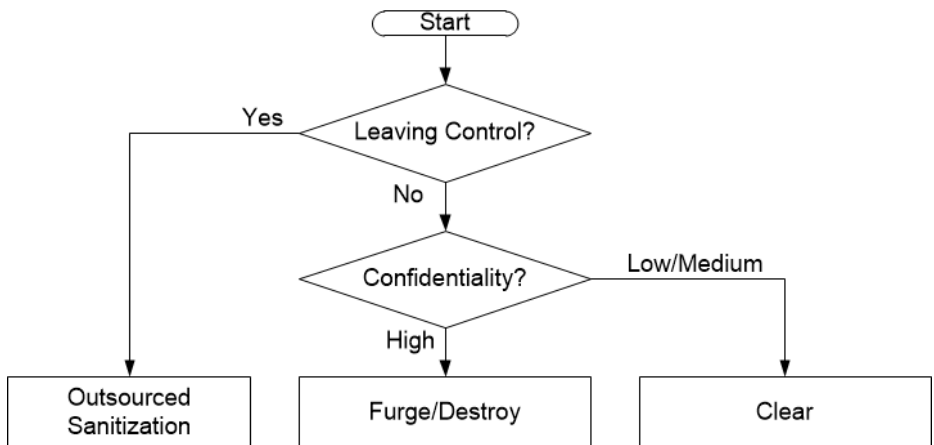
Figure 1. Media Sanitization Process is shown.

Conventionally, there is a guideline for destroying privacy data in NAND flash memory. As shown to FIG. 1, according to NIST standard, there is a description of flash memory according to Clear/Purge. As mentioned in the description of Clear in NIST standard, a flash memory-based storage device may not be able to completely delete old data even if it includes spare cells and performs wear leveling. This is because the storage device does not specify direct addressing of all areas where important data is stored. The clear operation here is to use software/hardware products to overwrite addressable space with non-sensitive data according to conventional read/write commands.

NIST standard also mentions the destruction of privacy data called 'Purge'. Here, purge operation includes overwriting, block deletion, and encryption deletion. However, since NAND flash memories do not normally support overwriting and is likely to store privacy data in a non-addressable space, it is difficult to accept that the NIST standard has appropriately performed privacy data destruction. Privacy data that is not mapped addresses still remains in NAND flash memories.

### A. Garbage Collection

In the case of a general flash memory, if a free block in the user block area is insufficient due to a bad block or the like, a free block in the spare block area is allocated as a user block area [12]. At this time, since the allocated size is block units (or two or more blocks), a larger space is allocated than the required storage space. Also, the free block in the spare block area becomes short in a relatively short time, and the flash conversion layer performs a background operation such as garbage collection in order to secure free blocks. Garbage collection is a process of selecting a specific block in a memory array, copying a valid page of a specific block to a free block, and then erasing the block to make it a free block, referring to FIG.2.

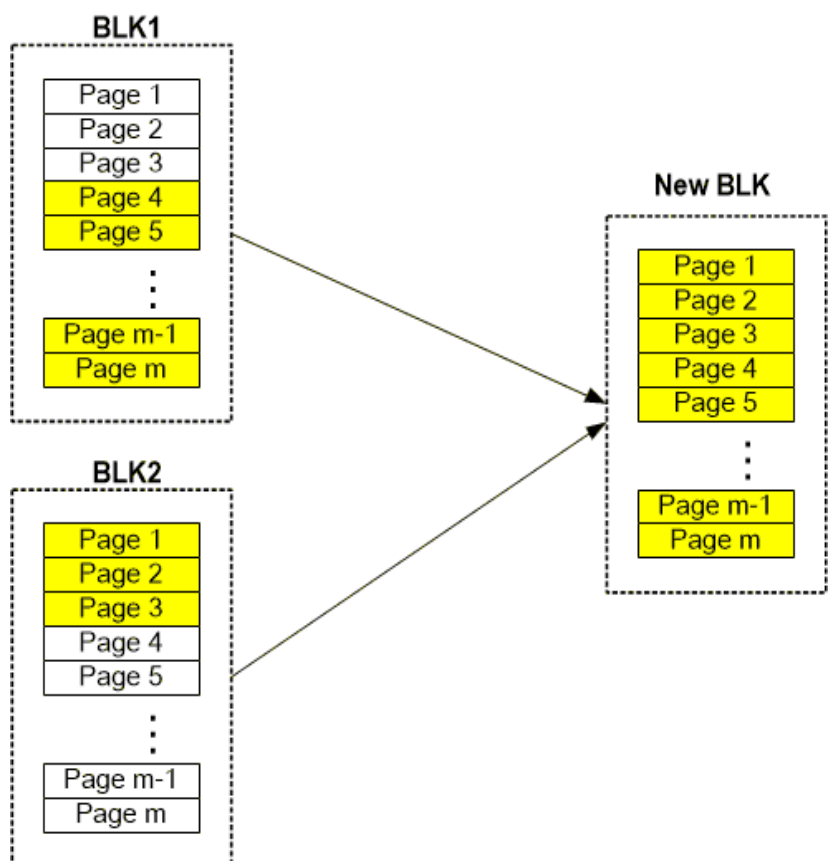

Figure 2. Garbage Collection for making a free block is shown. A garbage collection for collecting valid pages of the first block BLK1 and the second block BLK2 to generate a new block is shown.

The erased block may be used to record the data later. However, according to the internal policy, there is a problem that the erase operation for such a block does not progress rapidly. In this case, if the valid page includes privacy data, the invalid page may contain the same privacy data. Privacy data is contained in both the valid page in the free block and the invalid page in the block to be deleted. As a result, the privacy data is likely to be included in at least two blocks, only one of which is mapped and the others will remain in the unmapped state. As shown in FIG. 3, if conventional garbage collection is performed, privacy data that cannot be managed exists in at least two blocks. The unmapped first block BLK1 and the second block BLK2 retain their private information until the erase operation is performed.

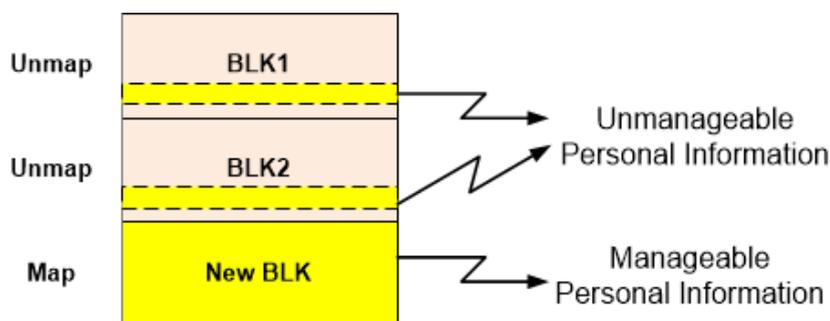

Figure 3. Unmanageable Privacy data is shown. Unmanageable privacy data remaining in the invalid blocks BLK1 and BLK2 after garbage collection is shown.

Our concern is how to destroy the privacy data contained in the invalid page in the block to be deleted. One of the simplest ways to destroy privacy data is to immediately erase blocks that contain invalid pages that contain privacy data. However, since the erase operation is performed on a block basis, it is difficult to manage the storage device in terms of policy. In general, one block is composed of 1024 pages. Performing an erase operation on one block due to an invalid page storing privacy data among 1024 pages is significant in management cost.

Existing NAND flash technology does not support overwriting. A reliable method of data invalidation is to perform an erase operation on BLK1 and BLK2. However, as is known, the erase operation is one of the management operations to be avoided as much as possible, since the number of erase cycles for the memory block is limited in terms of wear leveling. Furthermore, the erase operation consumes a large amount of power, and the erase time is also considerable. From the vendor's point of view, the focus is on keeping these erase operations from occurring as soon as possible.

Our research has begun to meet the needs of users who want to worry about these vendors and remove complete privacy data from flash memory. Ahn and Lee attempted to solve this problem by overwriting them in their studies [9]. To summarize their argument: The conventional method of writing NAND flash memory does not support overwriting, but since it is accessed from the viewpoint of privacy data destruction, it is intended to overwrite invalid pages with arbitrary data. Of course, any data here must be writeable data to the currently stored invalid page. In general, since the time required for one page write (program) operation and the erase operation for one block take 1000 times or more, this method is extremely effective as a privacy data destruction method.

### B. Data Modulation

In general, data is not stored in the form of original data in NAND flash memory. In order to minimize the possibility of data modification due to interference between data to be programmed, the original data is randomized according to predetermined rules and the randomized data is stored in a physical area [13]. As shown in FIG. 4, an encoder for modulating the data and a decoder for demodulating the data read in the physical area are essentially included.

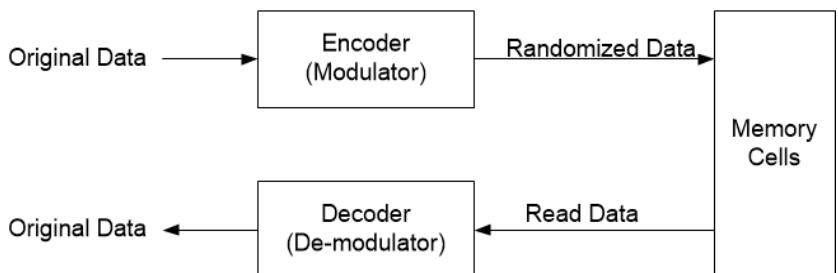

**Figure 4. Data modulation of NAND flash memory is shown.**

The program operation programs the modulated data into the memory cells after modulating the error null data. The read operation may output the original data after de-modulating the data read from the memory cells. The binary bits corresponding to the original data, that is, the private information, are not strictly stored in the memory cells. However, since the modulated data is stored in the memory cells, and the stored modulated data and the original data are mutually exchangeable by a predetermined method, the modulated data is also regarded as private information.

### C. State Mapping

The data modulated data is mapped according to the program state corresponding to the threshold voltage (Vt) of the memory cell.

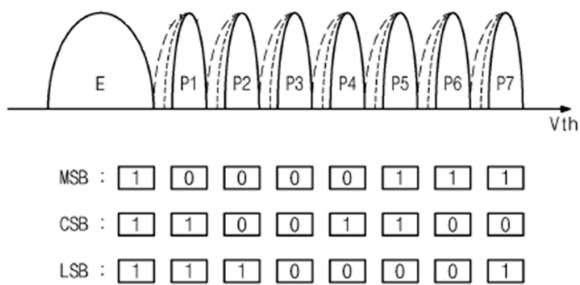

**Figure 5. State mapping in 3-level cell program is shown. States programmed into the memory cells in the three-level NAND flash memory and the corresponding data bits are shown.**

As shown in FIG. 5, for example, in the case of a 3-level cell, E state is mapped to '111', P1 state is mapped to '011', P2 state is mapped to '001', P3 state is mapped to '000', P4 state is mapped to '010', P5 state is mapped to '110', P6 state is mapped to '100', and P7 state is mapped to '101'. That is, program states with MSB (Most Significant Bit), CSB (Center Significant Bit), and LSB (Least Significant Bit) corresponding to the threshold voltage of the memory cell are determined [14].

Performing a write operation in a NAND flash memory means making each memory cell have any one of the eight threshold voltages. NAND flash memory basically includes a plurality of memory blocks, each of the plurality of memory blocks includes a plurality of pages, and each of the plurality of pages includes a plurality of memory cells. In general, NAND flash memory performs a write operation or a read operation in units of pages and performs an erase operation on a block-by-block basis.

## III. Possible Overwriting in NAND Flash Memory
Generally, it is known that NAND flash memories cannot be overwritten. However, there is at least one state in which NAND flash memory can be partially overwritten, and a data bit corresponding to the overwritable state exists. In 3-level programming, there is little probability that the programmed state will all become the top state P7. Since the randomization technique is applied, the probability that data corresponding to the highest state is written is 1/8 or less. This means that at least one of the states below the top state can be overwritten.

### A. Overwritable States
As noted in state mapping, each of the memory cells has a threshold voltage of any one of eight states. The write operation of NAND flash memory is repeated by sequentially applying a voltage to the word line corresponding to the page and verifying whether the threshold voltage associated with the corresponding bit is exceeded [9]. At this time, since the threshold voltage of the memory cell in which the program is completed is likely not to be the final state P7, states E to P6 lower than the final state P7 can be overwritten. What we are interested in is this overwritable state.

### B. Overwritable Data
As described above, it can be seen that there is a state in which overwriting is possible. Thus, if it is known which bit of data currently written in the memory cell is, the writable state is determined, and thus the overwritable data for the writable state can be determined [9].

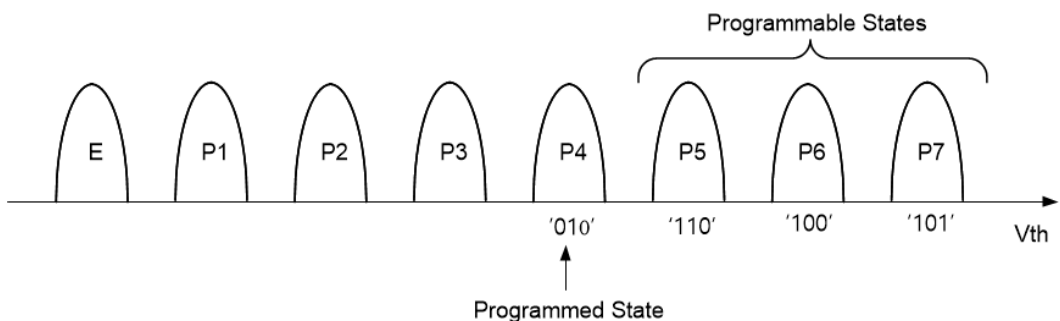

Figure 6. Overwritable Data are shown. Programmable states P5, P6, P7 are shown in a memory cell having a particular programmed state P4.

As shown in FIG.6, for example, if it is P4 state corresponding to '010' of the programmed memory cell, the overwritable state is the upper states, namely P5 to P7 states. Therefore, data that can be overwritten on the memory cell is '110', '100', and '101'.

In MLC program operations, it has been found that overwriting is possible for higher states than the programmed states. However, such overwriting is based on a program operation by default, which can lead to program disturbances. Here, the program disturbances affect valid data of neighboring pages by a current program operation. The deletion schemes proposed by us are designed to enable overwriting while minimizing the effects of these program disturbances.

In addition, in order to overwrite a page, it is inevitable to read the data of the current page, check the overwriteable state in the read data, and generate data for overwriting. However, if the costs for all of the processes described above are not greater than the cost of destroying personal information immediately in NAND flash memory, we will inevitably have to use this approach.

## IV. Proposed Privacy Data Destruction Schemes
We propose privacy data destruction schemes in NAND flash memory using overwritable states. The proposed privacy data destruction schemes include a partial overwriting scheme, a SLC programming scheme and a deletion duty pulse application scheme.

## A. Partial Overwriting Scheme

TABLE 1
PRIVACY DATA DESTRUCTION PROCESS USING PARTIAL OVERWRITE

| Personal Inform. | 661004 | | | | | | | | | | | | | | | |
|---|---|---|---|---|---|---|---|---|---|---|---|---|---|---|---|---|
| Binary Bits (48) | 001101100011011000110001001100000011000000110100 | | | | | | | | | | | | | | | |
| Random Bits (48) | 110110001101100011000100110000011000000011010000 | | | | | | | | | | | | | | | |
| State Mapping (16) | P5 | P5 | P2 | P7 | P6 | P1 | P3 | P6 | P5 | P3 | P2 | P6 | P3 | P1 | P4 | P3 |
| Partial Overwritable States (16) | | | | | | P2 | | | | | | | | P2 | | |
| | | P3 | | | | P3 | | | | P3 | | | | P3 | | |
| | | P4 | | | | P4 | P4 | | | P4 | | P4 | | P4 | | P4 |
| | | P5 | | | | P5 | P5 | | | P5 | | P5 | | P5 | P5 | P5 |
| | P6 | P6 | P6 | | P6 | P6 | P6 | | P6 | P6 | P6 | P6 | | P6 | P6 | P6 |
| | P7 | P7 | P7 | | P7 | P7 | P7 | P7 | P7 | P7 | P7 | P7 | P7 | P7 | P7 | P7 |
| Selected State (16) | P6 | P7 | P3 | P7 | P7 | P4 | P5 | P7 | P6 | P4 | P5 | P7 | P5 | P2 | P5 | P4 |
| PO Bits (48) | 100101000101101010110101100010110101110001110010 | | | | | | | | | | | | | | | |
| Derandom Bits (48) | 001100010011000000110000001100010011000000110001 | | | | | | | | | | | | | | | |
| Purge | 100101 | | | | | | | | | | | | | | | |

It is assumed that the invalid data of the memory block to be deleted contains privacy data. There will be binary data corresponding to the privacy data, and there will be data that modifies it. Information about the program status can be obtained through the respective status mapper of the modulated data. When program states corresponding to privacy data are acquired, data corresponding to states equal to or higher than the programmed states may be generated. And then a write operation can be performed on a page on which invalid data are programmed with such data. Privacy data may be discarded by such overwriting operation [9].

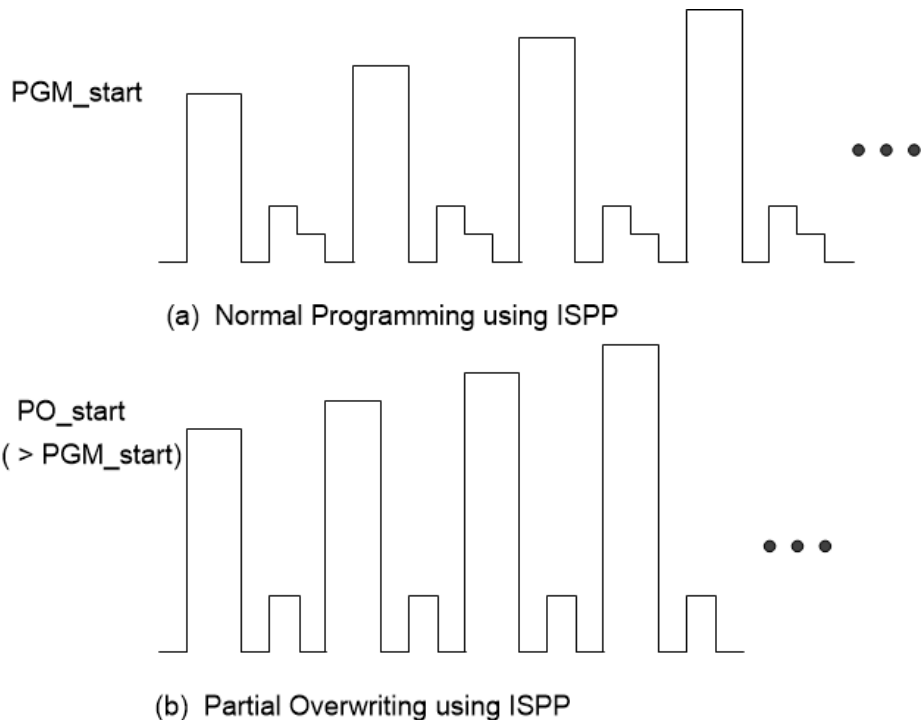

Figure 7. Partial overwriting using ISPP is shown. Note that (a) are pulses related to normal programming using ISPP, and (b) are pulses related to partial overwriting using ISPP.

In general, program operation of NAND flash memory is performed by ISPP (Incremental Step Pulse Programming) [15,16]. A specific pulse is applied to the word line and a pulse incremented by a predetermined value is applied to the word line in each step. This operation is repeated until state corresponds to the desired threshold voltage. After the initial program pulse (PGM_start), verify pulses are applied to verify state. When such a program set is advanced, a predetermined number of program pulses are applied, and then verify pulses are applied.

We propose to perform partial overwriting using ISPP described above. Of course, the initial overwrite pulse (PO_start) will be higher than the initial program pulse (PGM_pulse) of normal program operation. In addition, in the partial overwrite operation, the verify pulse can be set to less than the number of verify pulses of normal program operation. This is because the program state in which the overwrite operation is performed is expected to have a relatively high threshold voltage in most cases.

In the following, we will explain how to delete a personal dentification number using partial overwriting, for example. For convenience of explanation, it is assumed that the personal identification number is '661004'. The personal identification number can be changed in binary as follows. A state chain of Table 1 is selected in which a partial overwrite state is selected. Binary bits corresponding to '661004' are '001101100011011000110001001100000011000000110100'. By randomizing the binary bits through random algorithm, the randomized bits are '110110001101100011000100110000001100000011010000'.

States corresponding to the randomized bits are selected. The selected states are P5-P5-P2-P7-P6-P1-P3-P6-P5-P3-P2-P6-P3-P1-P4-P3. Each of the selected states is programmed in a corresponding memory cell of a page. As described above, program operations for states corresponding to privacy data are performed on the memory cells. After that, according to a request for destroying privacy data of the programmed memory cells, the partial overwriting may proceed as follows.

The partial overwritable states for personal identification number 661004 can be configured in any of a variety of overwrite enabled states, as shown in Table 1. In the topmost program state P7, no action will be taken. Therefore, the chain of program states selected for the partial overwriting is P6-P7-P3-P7-P7-P4-P5-P7-P6-P4-P5-P7-P5-P2-P5-P4. Thus, the selected state chain is programmed in the memory cells. State to be used for partial overwriting may be randomly selected.

TABLE 2
PRIVACY DATA DESTRUCTION PROCESS USING PARTIAL OVERWRITE

| Personal Inform. | 661004 | | | | | | | | | | | | | | | |
|---|---|---|---|---|---|---|---|---|---|---|---|---|---|---|---|---|
| Binary Bits (48) | 001101100011011000110001001100000011000000110100 | | | | | | | | | | | | | | | |
| Random Bits (48) | 110110001101100011000100110000001100000011010000 | | | | | | | | | | | | | | | |
| State Mapping (16) | P5 | P5 | P2 | P7 | P6 | P1 | P3 | P6 | P5 | P3 | P2 | P6 | P3 | P1 | P4 | P3 |
| Partial Overwritable States (16) | | | | | | P2 | | | | | | | | P2 | | |
| | | P3 | | | | P3 | | | | P3 | | | | P3 | | |
| | | P4 | | | | P4 | P4 | | | P4 | P4 | | | P4 | | P4 |
| | | P5 | | | | P5 | P5 | | | P5 | P5 | | | P5 | P5 | P5 |
| | P6 | P6 | P6 | | | P6 | P6 | | P6 | P6 | P6 | | P6 | P6 | P6 | P6 |
| | P7 | P7 | P7 | | P7 | P7 | P7 | P7 | P7 | P7 | P7 | P7 | P7 | P7 | P7 | P7 |
| Selected State (16) | P5 | P7 | P5 | P7 | P6 | P5 | P5 | P6 | P5 | P5 | P5 | P6 | P5 | P5 | P5 | P5 |
| PO Bits (48) | 100101000101101010110101100010110101110001110010 | | | | | | | | | | | | | | | |
| Derandom Bits(48) | 110001001100000011000100110001001100000011000100 | | | | | | | | | | | | | | | |
| Purge | ÄÀÄÄÄÄ | | | | | | | | | | | | | | | |

However, it is not necessarily limited to this, and may be selected as the highest state P7, or state P6 immediately below may be selected. When the highest state P7 is selected as state of partial overwriting, it should be noted that there is a possibility of physical destruction of the memory cell due to excessive program operation.

The partial overwrite scheme can verify that the destruction of privacy data was successful. When performing a state read operation from a partially overwritten memory cell, PO bits are '100101000101101010110101100010110101110001110010.' Derandomizing the PO bits, the derandomized bits are ' 001100010011000000110000001100010011000000110001.' Therefore, ASCII values corresponding to the derandomized bits are '100101'. Thus, the privacy data 661004 is completely destroyed in the memory cells.

On the other hand, in order to reduce the burden of selecting any one of the above overwritable states, the partial overwrite can proceed only to the program state lower than the specific program state. For example, the partial overwrite can advance the batch program operation to the fifth program P5 for states E, P1, P2, P3, and P4 lower than the fifth program state P5. That is, only the memory cells having a state lower than the fifth program state P5 are programmed to the fifth program state P5, and the memory cells having states higher than the fifth program state P5 maintain their current state.

A state chain of Table 2 is selected in which a partial overwrite state is selected beyond a specific program state. According to this method, the chain of program states selected for the partial overwriting is P5-P7-P5-P5-P5-P5-P5-P6-P5-P5-P5-P6-P5-P6-P5-P5-P5-P5 to be. PO bits corresponding to these states chain becomes '100101000101101010110101100010110101110001110010', and are programmed into the memory cells. ASCII values corresponding to this binary code is 'ÄÀÄÄÄÄ''. Therefore, privacy data is completely destroyed.

The above-described method is practically possible because it applies a randomization technique to equalize the number of bits to be programmed. When modulating the original data, the number under a specific state, for example, the intermediate program state, is similar to the remaining number because it is randomized. Therefore, the reverse analysis attack by the overwriting of this part by the attacker will be blocked as a source.

VOLUME XX, 2019                                                                                                                                             7

## B. SLC Programming

In general, NAND flash memory is performing multi-level cell program operation. When the target page is invalidated by an operation such as garbage collection, an on-the-fly invalid page program operation for destroying privacy data is performed before processing the corresponding memory block into an invalid block. Here, the on-the-fly invalidation page program operation constitutes a single-level cell program operation based on a specific reference level rather than a multi-level cell program operation.

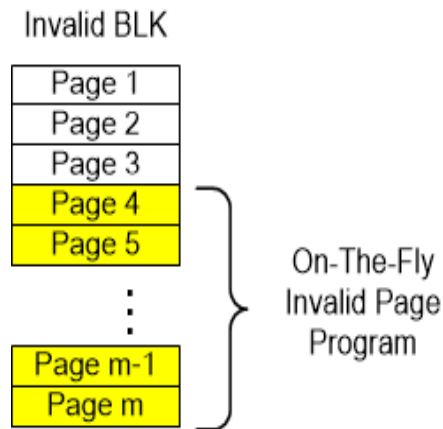

**Figure 8. On-the-fly invalid page program is shown. An on-the-fly invalid page program for a page with privacy data in an invalid block is shown.**

Our proposed on-the-fly invalid page program will perform a single-level cell program operation based on a certain threshold voltage, referring to FIG.8.

The memory controller can force an invalid page program on the fly on invalid pages prior to invalid block processing. This on-the-fly invalid page program operation can be achieved by performing a single-level cell program operation on random data in programmed multi-level cells. The concept is slightly different from the above-mentioned partial overwrite technique. In the partial overwriting technique, three pages of data are generated and programmed for privacy data destruction, but an on-the-fly invalid page programming technique is sufficient to generate and program only one page of data. In addition, there is no need to verify the success of the program.

As shown in FIG. 9, on-the-fly invalid page program operation using a SLC program (MLC PGM → SLC PGM) is shown.

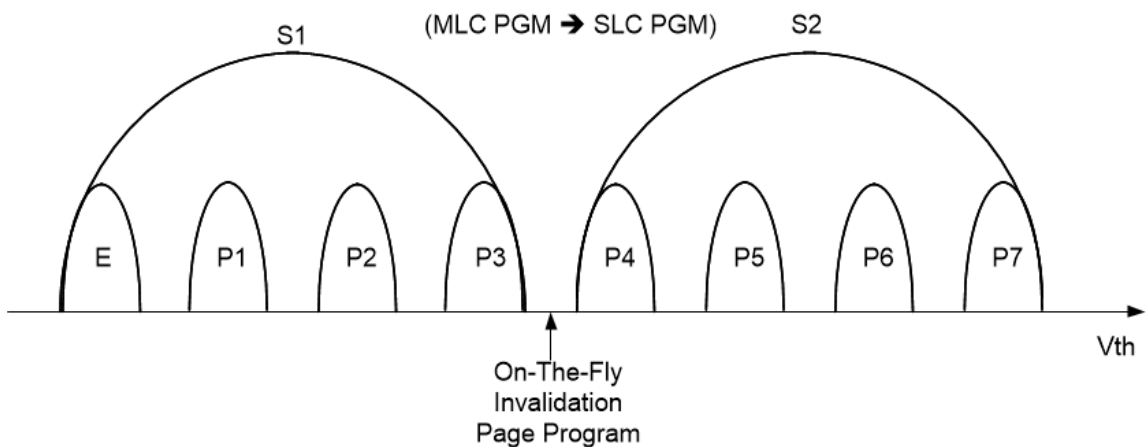

**Figure 9. On-the-fly invalid page program using SLC programming is shown. The on-the-fly invalid page program consists of a SLC program operation that performs a single-level cell program operation from a multi-level cell program.**

For example, the privacy data stored by the 3-level cell program is destroyed by the 1-level cell program, referring to FIG. 8. Therefore, it can be called the SLC programming technique.

## C. DDP Application Scheme

Generally, even when a power supply voltage is applied to a body region of a memory cell, if a word line voltage of a predetermined value or more is applied to a word line connected to the memory cell, charge transfer may occur due to F-N tunneling. This implies that the operation for invalid data destruction can be made surprisingly easily. That is, if the word line pulse is applied to the word line corresponding to the invalid page having the privacy data without applying the power supply voltage to the body side, the privacy data can be easily destroyed as a whole. This is because the threshold voltage can be changed into states that are completely unknown.

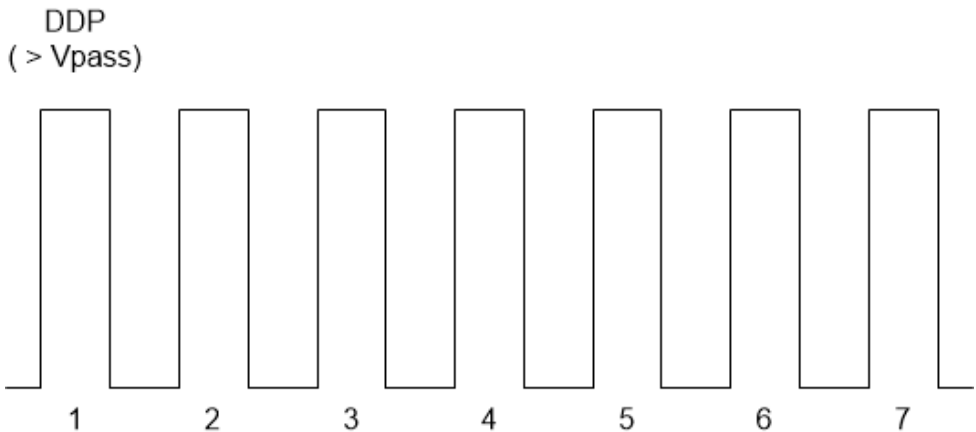

**Figure 10. Deletion duty pulse (DDP) is shown. Note that DDN pulse does not require a verify pulse and has the same pulse amplitude.**

These wordline pulses are called to be DDP (Deletion Duty Pulses), referring to FIG.10. By a request from the host or by NAND flash memory's own request, DDP can be applied to the invalid page. Since the data state is not required to be checked after such DDP is applied, a verify pulse is not required. The level of DDP may be higher than the pass voltage (Vpass). In general, when a voltage lower than the pass voltage (Vpass) is applied to the word line, there is no charge transfer in the channel. Also, the number of DDP should be determined experimentally. For example, the number of DDP may be determined smaller than the level or number of conventional program pulses. The number of DDPs can be determined to minimize program disturbances.

Generally, NAND flash memory uses ECC (Error Correction Code) technique for data correction due to deterioration of data. The number of erasure duty pulses must be determined to exceed the error correction level of this ECC technique. The number of such experimental DDP will depend on NAND chip manufacturer's cell characteristics.

*D. Request to destroy privacy data*

Upon receipt of the privacy data destruction request from the host, the memory controller preferentially identifies the block having the invalid page for the unaddressed area (e.g., BLK1, BLK2) in NAND flash memory. Thereafter, a partial overwriting (PO) operation or a delete duty pulse (DDP) operation is performed on a block having an invalid page.

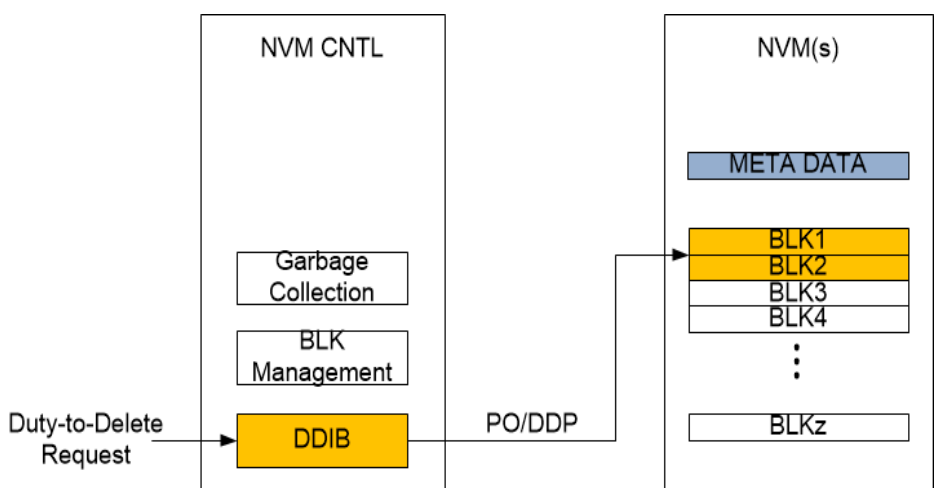

**Figure 11. Proposed storage device for processing Duty-to-Delete request is shown. The storage device may receive a privacy-related delete request from the host and may operate the privacy data destruction schemes in response to the request.**

As shown in FIG. 11, the memory controller may include a garbage collection unit, a block management unit, and a delete duty execution unit (DDIB). The block management unit can specially manage invalid blocks that have stored privacy data in comparison with the block management unit. For example, the block management unit may store the physical address for the invalidation page for the invalid block. The deletion obligatory execution unit may apply a privacy data destruction technique (e.g., PO / DDP) to the physical address for the invalidation page.

### E. Verification of privacy data destruction

In response to the host's privacy data destruction request, NAND flash memory can perform a private information destruction operation. Thereafter, NAND flash memory may perform a verification operation on the request for destroying privacy data. The verification operation of the privacy data destruction request determines whether the information to be restored of the privacy data requested to be destroyed or the information to be restored exists in the valid page or the invalid page of NAND flash memory. For example, it can be determined whether the data in which privacy data and privacy data are modulated exists in the invalidation area or the validation area of NAND flash memory. If there is no the privacy data or data corresponding to the privacy data in NAND flash memory, the privacy data destruction request is determined to be successful. After NAND flash memory transmits this fact in response to the host. Basically, in order to verify privacy data destruction, privacy data search for unmapped memory blocks as well as mapped memory blocks should be performed.

However, this verification is not as easy as it sounds. This is because the above-described destroy request verification operation is effective only when it is assumed that privacy data exists only in a non-volatile memory such as a NAND flash memory. Actually, there is a possibility that the privacy data exists in volatile memory (DRAM, SRAM, etc.) temporarily uploaded. In order to verify the existence of privacy data, there must be data to be a reference, and such standard data is also likely to be present in the volatile memory. Therefore, after the verification operation is completed, it is basically necessary to clean the volatile memory in the storage device having NAND flash memory. More researches are needed on this part.

## VI. Performance Comparison

We compared performances between the conventional Block Erase scheme, the partial overwriting, SLC programming, and DDP scheme. Here the performances include the degree of degradation and elapsed time of the memory cell until privacy data is destroyed in NAND flash memory. For the sake of convenience, we have compared the schemes to the garbage collection situation mentioned in section II.

Assume that there are M valid pages among the two memory blocks and N invalid pages. Here M valid pages will be programmed into a new memory block. Assume that there are totally N invalid pages in the two memory blocks. Thus, in all schemes, M program operations are performed on a new memory block by the garbage collection. In addition, the read operation for reading valid pages is not mentioned in the performance index because the degree of degradation is inferior to the erase operation or the program operation. here a is the degree of cell degradation due to the erase operation and b is the degree of cell degradation due to the program operation. In general, a is 1000 times higher than b [9, 17]. In other words, the degree of degradation due to the erase operation is more severe than that of the program operation.

TABLE 3
GARBAGE COLLECTION CASE

| Schemes | Degree of Cell Gradation (Memory Block) | | | Privacy data Destruction Time |
|---|---|---|---|---|
| | PGM count | Erase count | Total | |
| Block Erase Scheme | 0 | 1 | a | x (until GG, >M*$T_{PGM}$) |
| Lin's Scheme | x | x | x | x |
| Partial Overwriting Scheme | < N | 0 | < b*N | M*$T_{PGM}$ +N* $T_{RDG}$ +$T_{POW}$ |
| SLC Programming Scheme | < N | 0 | < b*N | M*$T_{PGM}$ + N*$T_{SDG}$ +$T_{SLCP}$ |
| DDP Scheme | < N | 0 | < b*N | M*$T_{PGM}$ +N* $T_{DDP}$ |

In the case of a block erase scheme, no program operation is performed on the two memory blocks to be invalidated. Instead, a program operation is performed on M valid pages for a new memory block. Therefore, in the block to be invalidated, the erasing operation is performed once to delete privacy data. Typically, block erase operations in garbage collection are performed in accordance with NAND flash memory's internal policy. This internal policy is managed to maximize the life of memory blocks. Therefore, as shown in Table 3, privacy data destruction time is virtually impossible to predict by various operating variables.

On the other hand, the proposed schemes can predict the destruction time of privacy data by applying the deletion schemes. For example, time of the partial overwriting scheme is M*$T_{PGM}$+N*$T_{RDG}$+$T_{POW}$, time of SLC programming scheme is M*$T_{PGM}$+ N*$T_{SDG}$+$T_{SLCP}$, and time of DDP scheme is M*$T_{PGM}$+N*$T_{DDP}$. Where $T_{PGM}$ is the program operation time, $T_{RDG}$ is random data generation time, $T_{SDG}$ is SLC data generation time, T is the overwriting time, $T_{POW}$ is the SLC programming time, and $T_{DDP}$ is the DDP application time.

VOLUME XX, 2019                                                                                                                                                    10

In summary, the schemes we propose can predict the destruction time of privacy data. In addition, comparing the degree of cell degradation for one memory block, the block erase scheme has a degree of degradation corresponding to one erase operation.

Lin's approach only considers the update situation, not the garbage collection situation at all. This suggests that, according to Lin's approach, personal information is still in the invalidation blocks in a garbage collection situation.

Also, we compared the performance of the deletion schemes assuming a case where an update is made to one page of privacy data in one memory block. On the other hand, Lin's scheme and our proposed schemes do not need to perform an erase operation until the privacy data is destroyed. Therefore, until the destruction of privacy data, our schemes have a degree of degradation that has performed less than N programming operations.

Even a single page update, the conventional block erase scheme cannot predict the destruction time of privacy data. On the other hand, all of Lin's scheme and our schemes can predict the destruction time. And the degradation characteristic up to privacy data destruction is also better in the case of the proposed schemes than the conventional erase scheme. As shown in Table 4, for example, time of Lin's scheme is $M*T_{PGM}+T_{ONESHOT}$, time of the partial overwriting scheme is $M*T_{PGM}+T_{RDG}+T_{POW}$, time of SLC programming scheme is $M*T_{PGM}+T_{SDG}+T_{SLCP}$, and time of DDP scheme is $M*T_{PGM}+T_{DDP}$. Where $T_{ONESHOT}$ is the one shot programming time.

TABLE 4
ONE PAGE UPDATE CASE

| Schemes | Degree of Cell Gradation (Memory Block) | | | Privacy data Destruction Time |
|---|---|---|---|---|
| | PGM count | Erase count | Total | |
| Block Erase Scheme | 0 | 1 | a | x (until GG) |
| Lin's Scheme | 1 | 0 | b | $M*T_{PGM}+T_{ONESHOT}$ |
| Partial Overwriting Scheme | 1 | 0 | b | $M*T_{PGM} + T_{RDG} +T_{POW}$ |
| SLC Programming Scheme | 1 | 0 | b | $M*T_{PGM} + T_{SDG} +T_{SLCP}$ |
| DDP Scheme | 1 | 0 | b | $M*T_{PGM} + T_{DDP}$ |

We confirmed that the proposed schemes can significantly reduce the destruction time of privacy data compared to the conventional block erase scheme. In addition, we can see that the proposed schemes have many advantages in terms of degradation characteristics of memory cells. As such, our proposed method of destroying privacy data can be easily applied in existing NAND flash memory, and seems to be useful in terms of management and economics.

However, in the future, the study of program disturbance for valid pages according to partial overwriting, SLC programming, and DDP application should be conducted in depth. We proceeded ignoring the read count, but we should also consider the factor of the read count in order to actually apply the delete obligation. The problem of disturbance due to the read operation is also discussed later.

## VII. Conclusion

We have identified privacy data that is bound to remain in NAND flash memory. Such residual privacy data is increasing the risk of disclosure of privacy data by illegal acquisition or unauthorized access of NAND flash memory. We discussed how to remove this residual privacy data from NAND flash memory. For example, we have proposed a way to identify personalized information by partially identifying programmable states and using partial overwriting schemes. We have also proposed a method of discarding privacy data by applying the delete duty pulse several times. We also proposed a method to destroy privacy data using SLC programming. Our private data destruction schemes in this NAND flash memory greatly reduce the memory performance degradation compared to the conventional erase operation and expect a considerable effect in terms of time and cost. Our proposed NAND flash memory privacy schemes are expected to have very important and positive consequences for future societies that have the duty to delete. In privacy technology, we are confident that our proposed techniques will be useful as anti-forensic technology of NAND flash memories.

**ACKNOWLEDGMENT**

We are appreciated to a professor Kwon for introducing us to the right to be forgotten.**REFERENCES**

[1] Y. Xu and Z. Hou, "NVM-Assisted Non-redundant Logging for Android Systems," 2016 IEEE Trustcom/BigDataSE/ISPA, Tianjin, pp. 1427-1433, 2016.

[2] Rize Jin and Tae-Sun Chung, "Dynamic regulation of index implementation for flash memory storages," 2010 The 2nd International Conference on Computer and Automation Engineering (ICCAE), Singapore, pp. 325-328, 2010.

[3] Q. Zhang, X. Li, L. Wang, T. Zhang, Y. Wang and Z. Shao, "Optimizing deterministic garbage collection in NAND flash storage systems," 21st IEEE Real-Time and Embedded Technology and Applications Symposium, Seattle, WA, pp. 14-23, 2015.

VOLUME XX, 2019	12[4] D. Chang, W. Lin and H. Chen, "FastRead: Improving Read Performance for Multilevel-Cell Flash Memory," in IEEE Transactions on Very Large Scale Integration (VLSI) Systems, vol. 24, no. 9, pp. 2998-3002, Sept. 2016.

[5] C. Matsui, A. Arakawa, C. Sun and K. Takeuchi, "Write Order-Based Garbage Collection Scheme for an LBA Scrambler Integrated SSD," IEEE Transactions on Very Large Scale Integration (VLSI) Systems, vol. 25, no. 2, pp. 510-519, Feb. 2017.

[6] J. P. van Zandwijk and A. Fukami, "NAND Flash Memory Forensic Analysis and the Growing Challenge of Bit Errors," IEEE Security & Privacy, vol. 15, no. 6, pp. 82-87, November/December 2017.

[7] A. Jonesm, O. Angelopoulo and L. Noriega, "Survey of data remaining on second hand memory cards in the UK," Computers & Security, ISSN: 0167-4048, Vol: 84, pp. 239-243, 2019.

[8] https://www.eugogo.eu/data-privacy

[9] N .Y. Ahn and D.H. Lee, "Duty to Delete on Non-Volatile Memory," https://arxiv.org/abs/1707.02842 , July 2017.

[10] P.H. Lin, Y.M. Chang, Y.C. Li, W.C. Wang, C.C. Ho, and Y.H. Chang, "Achieving fast sanitization with zero live data copy for MLC flash memory," Proceedings of the International Conference on Computer-Aided Design (ICCAD '18). ACM, New York, NY, USA, 2018.

[11] R. Kissel, A. Regenscheid, M. Scholl and K. Stine, "Guidelines for Media Sanitization," NIST Special Publication 800-88, Revision 1, 2014

[12] J. Gu, C. Wu and J. Li, "HOTIS: A Hot Data Identification Scheme to Optimize Garbage Collection of SSDs," 2017 IEEE International Symposium on Parallel and Distributed Processing with Applications and 2017 IEEE International Conference on Ubiquitous Computing and Communications (ISPA/IUCC), Guangzhou, 2017, pp. 331-3317.

[13] J. Cha, W. Kang, J. Chung, K. Park and S. Kang, "A New Accelerated Endurance Test for Terabit NAND Flash Memory Using Interference Effect," IEEE Transactions on Semiconductor Manufacturing, vol. 28, no. 3, pp. 399-407, Aug. 2015.

[14] M. Kang, W. Lee and S. Kim, "Subpage-Aware Solid State Drive for Improving Lifetime and Performance," IEEE Transactions on Computers, vol. 67, no. 10, pp. 1492-1505, 1 Oct. 2018.

[15] P. Wang et al., "Three-Dimensional nand Flash for Vector–Matrix Multiplication," IEEE Transactions on Very Large Scale Integration (VLSI) Systems, vol. 27, no. 4, pp. 988-991, April 2019.

[16] J. Ko et al., "Variation-Tolerant WL Driving Scheme for High-Capacity NAND Flash Memory," IEEE Transactions on Very Large Scale Integration (VLSI) Systems, vol. 27, no. 8, pp. 1828-1839, Aug. 2019.
T. Chen, Y. Chang, C. Ho and S. Chen, "Enabling Sub-blocks Erase Management to Boost the Performance of 3D NAND Flash Memory," 53nd ACM/EDAC/IEEE Design Automation Conference (DAC), Austin, 2016.